\documentclass[preprints,article,accept,moreauthors]{Definitions/mdpi} 

\usepackage{rotating} 
\usepackage{graphicx}

\firstpage{1} 
\pubvolume{1}
\issuenum{1}
\articlenumber{0}
\pubyear{2026}
\copyrightyear{2026}
\datereceived{9 October 2026} 
\daterevised{13 October 2026} 
\dateaccepted{ } 
\datepublished{ } 

\Title{Detectability of Spooky Quasar-Flux Correlation on the Sky}

\Author{Eric Steinbring $^{1}$\orcidA{}*}

\AuthorNames{Eric Steinbring}

\address[1]{%
$^{1}$ \quad National Research Council Canada, Herzberg Astronomy and Astrophysics, Victoria, BC V9E 2E7, Canada; Eric.Steinbring@nrc-cnrc.gc.ca}

\corres{Correspondence: Eric.Steinbring@nrc-cnrc.gc.ca}

\abstract{The fluxes from two quasars, picked at random, are expected to be found uncorrelated. In fact, high-speed photometry of an antipodal pair can source acausal photons that might serve as ideal inputs to a quantum-mechanical experiment. Could less-separated quasars be found correlated? This question motivates a survey on timescales of minutes to within a night or longer, for which years of suitable archival data already exist. A potential signal is envisioned by applying the holographic principle, requiring cosmic isotropy, and avoiding bias due to photometric errors between any two samples. This restricts detectable quasar flux-differences from an overall average, leaving randomness intact over the whole sky on long timescales as required, but with a prominent peak: those near that angular separation are predicted to more-often-than-not share a slight but measurable excess brightening/dimming. Analysis of one dataset from the Gemini twin telescopes is presented, using over 14 years of archival broadband-optical GMOS North and South images, serendipitously sampling thousands of quasars in g, r, i and z bands. The model fits the data well, rejecting a null result of uniform randomness with 97.7\% confidence, instead consistent with a $3\sigma$ residual correlated signal of 0.21 mag peaked at $65^{\circ}\pm3^{\circ}$ separation. Confirmatory observations with future wide-field galaxy and quasar surveys are discussed.}


\begin{document}


\section{Introduction}\label{introduction}

Non-correlation of fluxes between two widely-separated quasars can be utilized for quantum-mechanical (QM) experiments. Independent measurements of each, made with two different receiver telescopes, should ensure that those two photon streams are also demonstrably independent, uncorrelated by more than mere random chance. This allows a fair, unbiased switching mechanism to a QM test \cite{Clauser1969}, triggering based on whether, say, a red or blue photon happens to be detected at either \cite{Friedman2013, Gallicchio2014}. Ultimately, for an antipodal pair (separated by exactly $180^{\circ}$) both of $z\geq3.65$ those photons are acausal, so they could never have communicated nor be physically synchronized. They should act as an ideal random-number generator \cite{Rauch2018, Steinbring2025}. Thus, a compelling reason to compare flux measurements between any two lower-$z$, less-separated quasars would be to investigate if, for those, there could be some measurable correlation between photons when sources are causal. One potential form of such correlated fluxes follows via the holographic principle \cite{t'Hooft1993}; recognizing that quasar fluxes might act like quantum states of a system, and integrating their total averaged differences over all quasars (which should be zero), while satisfying both the required acausality condition (for just those separated by $180^{\circ}$) and cosmic isotropy (globally, among all sources) simultaneously can lead to an expected, asymmetric "residual" distribution of sampled flux-differences (brightening of one, dimming of another) on the sky. This study will describe that potential residual signal, and look for it in photometry of quasars.

Quasars are relatively bright and compact, far exceeding flux from their host galaxy. Although variable on all timescales \cite{Mudd2018}, even a typical $z\approx4$ quasar in the optical is $20~{\rm mag}$, well within reach of an 8-m class optical/near-infrared telescope with modern CCD imagers to pointsource signal-to-noise ratio greater than 5 at modest exposures of about 60~{\rm s}. Many facilities have been (serendipitously) gathering flux measurements of those and brighter sources nightly for decades. Much of these data are public, and simple aperture photometry is sufficient to perform the measurements and calculate the differences between between any two on all of those timescales; so a survey looking for previously unsensed correlation should be possible. Do far-separated quasars brighten and dim in unison?

A bias to such a search from the ground for separation angles greater than roughly $90^{\circ}$ on timescales less than about 12 hours is, of course, that the Earth gets in the way. From the nightside, any single optical/near-infrared observatory sampling the sky on less than an hour timescale, is restricted to workable separations below about two airmasses. At that separation, these incur at least twice the zenith extinction for each, even under photometric skies, with similarly degraded seeing. Darkness reaches only $\sim 21~{\rm mag}~{\rm arcsec}^{-2}$ in the visible, which is relevant as fluxes grow faint, by worsening their colour-discrimination. That matters, as those fluxes (in ${\rm photons}~{\rm s}^{-1}$) signal the switch settings in practice in the aforementioned QM tests, and when their relative photometry (${\rm photons}~{\rm s}^{-1}~{\rm arcsec}^{-2}$) is ill constrained these choices may be subject to an unsensed bias. Tolerances to this are as little as 14\% \cite{Friedman2019}. But that is a readily testable photometric limit for exposures of about 1 minute or so, and a method of sensing average flux differences will be described here, for conditions where point-source photometry reaches those limits.

It happens that there is at least one ideal dataset to probe. Gemini, on Maunakea in Hawaii ($19.82^{\circ}$N, $155.47^{\circ}$W, 4213 m) and on Cerro Pachon in Chile ($30.24^{\circ}$S, $70.74^{\circ}$W, 2722 m), when each viewing a target near zenith, places those $95.5^{\circ}$ apart on the sky; so it naturally provides unbiased sampling of separation angles up to the full sky on all timescales from minutes to within a "night" of about 17 hours. These 8-m class telescopes have operated identical optical imagers continuously for about two decades; the Gemini Multi-Object Spectrograph North and South (GMOS-N, GMOS-S) and a public archive eases aggregation of many serendipitous observations of quasars; typically with exposures of about $60~{\rm s}$. This consists of photometry in ${\rm g}$, ${\rm r}$, ${\rm i}$ and ${\rm z}$ filter bands for thousands of quasars with redshifts $0.1<z<6$, sensitive down to 23 mag. Cross-comparing all quasar photometry and calibration-star data taken together within those frames across this timeframe can create exactly the unbiased dataset needed to look for flux-correlation between them on timescales of minutes to many hours, and longer.

The next section presents a gedankenexperiment: a QM test involving all-possible quasar pairs, leading to a potential form of correlated quasar fluxes with angular separation, especially prominent at $64.6^{\circ}$. Correlated photometric errors inherent in photometry of widely-separated pairs in a serendipitous survey from Gemini are considered, but do not harm this expectation. Following that, the available Gemini dataset is described. The final sample comprises roughly 2 million observational pairings, which in their aggregate (0.25-mag $1\sigma$ uncertainty within 6-degree-wide sampling bins) is sufficient to show a biased difference in brightness between quasars relative to object separation, reaching a peak signal consistent with the model expectation and against a null hypothesis of flat. Although this suggests an intriguing connection between those quasar fluxes, it is not in conflict with either causality or previous QM results, although invites further investigation. Follow-up targets of interest are provided. Discussion concludes on prospects of wide-field surveys such as with Rubin, and reaching necessary photometric accuracy to verify this result.

\section{Methodology}

The approach followed here is to envision an angular dependence of correlation between quasars that could occur, but has not been looked for previously. Inspiration comes from a cosmic application of the holographic principle, as has been used elsewhere to look for signatures of quantum gravity due to blurring-out of pointsources \cite{Ng2025, Steinbring2026}. This central tenet of information-theory prescribes that integrating over the maximal number of quantum states within a volume - in this case, all possible absolute instantaneous flux-differences between any two observable quasars in the Universe - should reduce to a property measurable over its containing surface as, say, a distribution of brightened/dimmed quasars over the whole sky. This calculation will be found to add to their dispersion in brightnesses for a particular separation, and a survey will be described, using Gemini to look for that effect.

Quasars can be assumed isotropic about Earth, and their total luminosity summed over all time is fixed, if unknown. Imagine that, in ensemble, all quasars act together as components of one universal QM experiment, an all-encompassing version of what is mentioned in the introduction. The different periods of activity of active-galactic nucleii (AGN) central engines, their redshift and luminosity distributions, or other properties are immaterial, as those are (in this scenario) all simultaneously viewed. And if so, then the outcome of a test performed with any two of those must be {\it perfectly} predictable, as their net correlation can be exactly known, if in total those must sum to a constant. The exact internal details of an actual QM-test apparatus itself are not critical, but generically, it depends on a setting-angle within it. That arrangement is now described in order to make a prediction of what correlation between any two quasars may be, simply by hypothesizing that those angles can be linked together by requiring this cosmic QM test avoids anisotropy.

\subsection{The Discriminator Angle in a Bell-Test Arrangement}

\begin{figure}
\center{\includegraphics[width=10 cm]{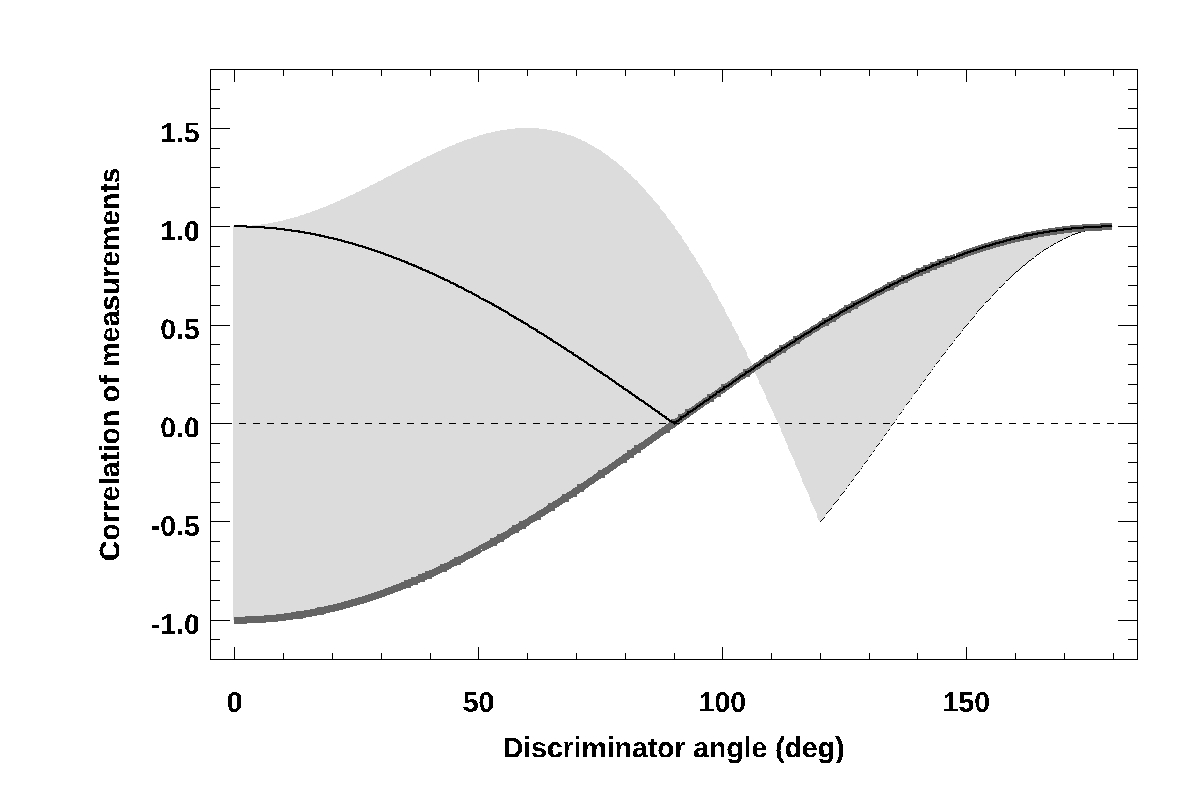}}
\vspace{-0.25 cm}
\caption{Quantum-state correlation by discriminator-angle in an idealized Bell-theorem experiment, for the expected QM result (thick-black curve) or its absolute value (thin-black curve) against the maximal difference between these, and that requiring hidden variables (shaded light grey).}\label{figure_bell_correlation}
\end{figure}

Quantum theory demands that entangled photon or electron pairs must be found in opposite states; if one is found with horizontal polarization or spin-down, the state of its entangled twin will {\it always} be found vertical or spin-up. Without recourse to hidden variables (or superluminal signals) this has been dubbed "spooky action at a distance" \cite{Einstein1935}. Importantly, any real experiment cannot measure both states in one direction simultaneously, as this requires a setting change. For example, a test which employs polarimetry necessitates a discriminator, such as a polarizer or the rotation of a waveplate. A choice must be made as to which polarization angle (or arm) to sample. Detecting one entangled state instantly collapses the wavefunction of both subject to shared uncertainty, with a probability density $q$ which depends on the angle $\theta$ between waveplate settings (see Figure~\ref{figure_bell_correlation}). Those states must be anti-correlated when co-incident (there are exactly two possibilities) and preservation of equal average probabilities of both states implies no net correlation, and so a functional form for normalized correlation of $-\cos{\theta}$, crossing at $\pi/2$. But when discriminator settings are truly random, this can be fairly compared to the form \citep{Bell1964}: $$p(\theta) = |\cos{2\theta} - \cos{\theta}| + \cos{\theta},\eqno(1)$$which exceeds unity at all lesser angles. Looking for this "excess" correlation is the test, as the difference $D$ between the two curves discriminates between QM and mimicry by hidden-variable theories; at, say, $0^\circ$ and $180^\circ$. (A better practical selection is not important here.) And in a test using quasars this binary polarizer/waveplate measurement-position choice is set randomly via their red and blue photons at the two receiver telescopes.

To be clear: the two (entangled) photons within the experimental setup are not the same two external photons that arrive from distant quasars to randomly set the discriminator angle. It would be only a coincidence that the QM test apparatus itself also involved a binary measurement of photon states. Actually, in the original theoretic description (and for some real tests) electrons are employed, detecting either spin down or spin up. But it will be shown that a direct connection can be made between the arrival angle of the external triggering photons at Earth and the internal discriminator-angle dependent outcome.

\subsection{Potential Correlations Across the Sky}

In a real experiment there is no influence on results caused by correlation between the two fair external sources and choice of discriminator angle. These are random, without bias in their triggering property: any red or blue photon (within suitably defined passbands, corrected for relative colour) from one quasar will correspond just as likely as not with receiving the opposite from the other. But what if that is not true? A scenario is described in Appendix~\ref{appendix}, connecting the probability of a measureable difference in outcomes via the trace-distance metric \citep{Friedman2019}$$P={1\over{2}}+{D(p,q)\over{4}},\eqno(2)$$with a 50\% chance of either $D=|p-q|=0$ or 2 for perfectly random switches (Case:~\ref{appendix1}). But correlation {\it is possible} for causally-connected sources, influencing the outcome, as when external differences in red versus blue are neutral for only one or less of every 4 photons, so non-neutral for every 3 or more. And presumably some "extra" correlation $M$ could be reached at which an effect on $D$ cannot be hidden, before being necessarily spoiled by full mimicry of QM. If so, there exists two {\it perfectly correlated} quasars which always share red for blue photons giving the opposite trigger to every photon arriving at either receiver, and (absent measurement error) {\it always} setting the angle to acheive an outcome as if they were now internal to the apparatus (Case:~\ref{appendix2}). The quasars effectively share "states" red/blue equivalent to the internal horizontal/vertical-polarizations (or spin-down/spin-up) and are also necessarily entangled in exactly the same way. Importantly, this also demands a single preferred orientation of the receivers on the sky for the test.

If such perfect sources could be found it would clearly violate cosmic isotropy, even if those quasars are not antipodal as viewed from Earth. And, furthermore, in an imagined scenario where all quasars are monitored, any spatial orientation can become favoured, available to foil the test. In that case the two receiver telescopes can {\it randomly} point to a source for which they need never know the other's internal apparatus settings, just some sky position in equatorial right ascension $\rho$ and declination $\delta$. In this case, measurement $D$ depends {\it solely} on the angle $\phi$ between sources, that is, $q(\rho, \delta)\leftrightarrow q(\phi)$ and similar for $p$ with the same functional form but no $\theta$ dependence. But consider relaxing the ability to detect perfect correlation, i.e. a red photon incident at $180^{\circ}$ from a blue one triggers results as in Figure~\ref{figure_bell_correlation} half the time. When the influence of $\phi$ dependence falls below, that difference is instead\footnote{By adding up the 12 unique terms thus formed with $\pm M$, $(D/2)^2=(p-q+M)(p+q-M)+(q-p+M)(q+p+M) +$ ..., all the cross terms cancel.} $$D(\phi) = \sqrt{M^2+p^2(\phi)-q^2(\phi)} ~; ~0.75 \leq M \leq 1,\eqno(3)$$so it suggests locating a "special" non-neutral pair (usually correlated) could still trigger results to correspond more often than not. This motivates the definition of a relaxed metric $$D_R/2 \equiv 3[D(\phi) - R]/4,\eqno(4)$$anticipating how $R$ could manifest as a shared, measurable residual correlation through cross-comparison of pairs, but avoiding the non-physical anisotropic case. 

\begin{figure}
\hspace{-1cm}
\includegraphics[width=10 cm]{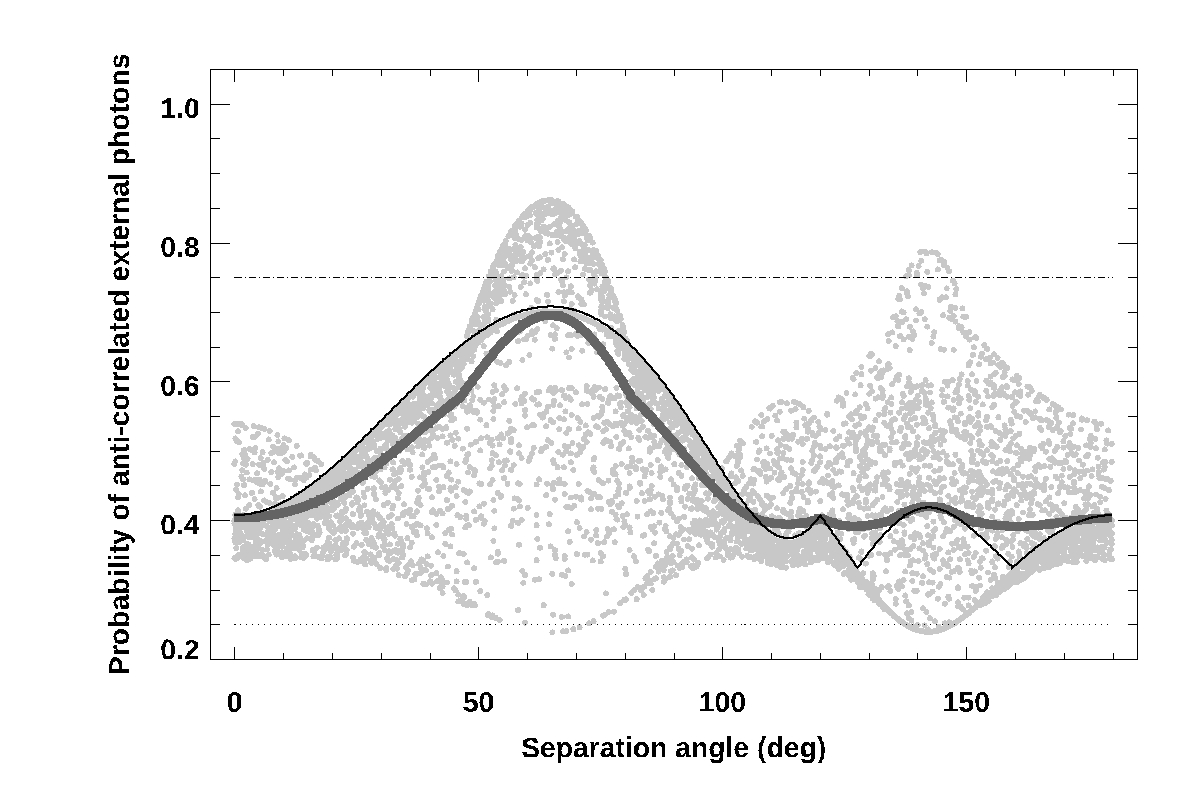}\includegraphics[width=6.7 cm]{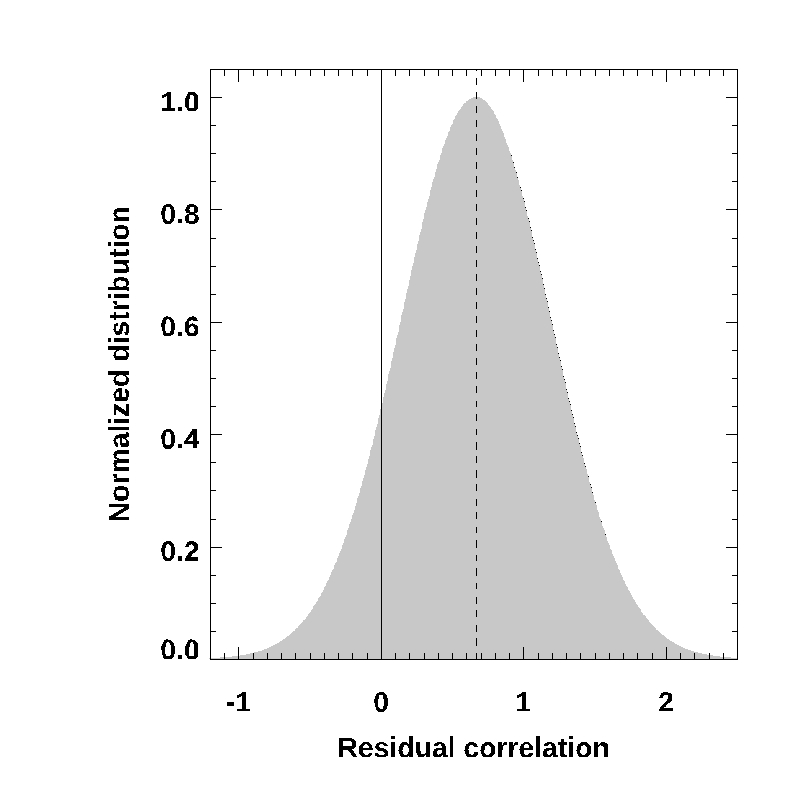}
\caption{Left panel: minimal absolute bound for the probability of flux-correlation as a function of quasar separation (thin black curve). Light-grey points show simulated output of 10000 samples discussed in \S 2.3, with $\bar{R}=0.67$, giving the probabilities by separation of finding two giving perfectly matched red-for-blue photons; maximally to $M=0.87$ against global neutrality ($\bar{P}=50\%$) preserving insufficient correlation to spoil a QM-experiment, within thresholds of 25\% and 75\% (horizontal dotted lines). Pairings above the last would belong to correlation-biased cases: just slightly more likely than not to be red when the other was blue, or vice-versa. The overall average of all trials by separation is given by the thick grey curve. Right panel: distribution of $R$ in a full-sky search.}\label{figure_probability}
\end{figure}

Nature might provide these special, but imperfectly-correlated quasars without violating cosmic isotropy or altering the internal mechanics of the test, and an application of the holographic principle puts limits on finding those, as one could seek, via their average correlations, experimental pairings simply {\it more likely} to provide correlated photons. To find them, take a complete census of every possible pairing, that is, the sum-total of all instantaneous red/blue quasar states within the volume of the Universe:$$\bar{P}=\int\displaylimits_{\rm Vol.}d{P}=\int\displaylimits_{\rm R.A.}d{\rho}\int\displaylimits_{\rm Dec.}d{\delta}{P}={1\over{2}} + {\overline{D_R}\over{2}}~~~~~~~~~~~~~~~~~~~~~~~~~~~~~~~~~~~~~~~~~\eqno(5)$$$$~~~~~~~~~~~~~~~~~~~~~~~~~~~~~~~~~~~~~={1\over{2}} + {3\over{4\pi}}\int\displaylimits_0^{\pi} d\phi[D(\phi) - R],$$tallying by each $\delta$ in polar coordinates about azimuth $\rho$, or equivalently the projected great-circle angular distance $\phi$ between any two sources. This requires only that each is uniquely identifiable by position on the sky, and so those cannot be perfectly overlapping. It is reasonable for near-pointsources, even though not strictly true for quasars, as they are extended. Isotropy demands $\bar{P}=50\%$ (all sky directions equal, with no preferred orientation) thus, by inspection, the intregral in equation 5 must be within the lower and upper possible limits of equation 4 for $R=\overline{|D_R|}=0$, presenting the desired tool: $$D_0\equiv(|D_R|-R)/2~~~~~~~~~~~~~~~~~~~~~~~~~~~~~~~~~~~~~~~~~~~~~~~~~~~~~~~~~~~~~~\eqno(6)$$$$~~~~~~~~~={3\over{4}}\sqrt{1 + p^2(\phi) - q^2(\phi)}-0.882989... ~;~ M=1, R=0,$$the constant arrived at by numerical integration of the left-hand term. Adopting instead $M=\sqrt{3/4}\approx0.87$, fairness in equation 2 is still not violated at antipodes (and is plotted as a thin black curve in Figure~\ref{figure_probability}) but continuing to bound $D$ by $-2R$ ($P=25\%$; resulting in 1 of 4 switching photons corrupted) everywhere, it has an upper limit $|D_R|+R$, defining an allowed lower envelope to any detectable correlation: $$P(\phi)={1\over{2}} + {D_0\over{2}}, ~~~~~~~~~~~~\eqno(7)$$where $|D_R|/2 \leq R$, which is counterbalanced above by $$P(\phi)={1\over{2}} + 2 D_0 - {3\over{4}} R,\eqno(8)$$that is, a probability distribution with a peak restricted to $M$, and revealing those pairs which offer more-often-than-not correlated photons: above the $P=75\%$ threshold.

Evidently, $R$ measures some detectable amount of correlation between any two quasars. That need not be a constant, and an attractive solution is arrived at by assuming this is instead distributed completely at random among them. If so, equation 5 perfectly balances near $\bar{R}=2/3\approx 0.67$, with the third term in equation 8 vanishing, leaving $2D_0$. Notice a maximum occurs at $\phi=64.6^\circ$, which is different than $p$, at $60^\circ$. Interestingly, a value of $2/3$ is the same as the so-called alpha-exponent in holographic quantum-foam models, which limit detectable spatial information, as has been seen with the blurring of quasars and gamma-ray bursts \cite{Steinbring2007, Steinbring2015, Steinbring2023, Steinbring2025b}. And, in the same sense, $R$ also limits the intrinsic correlation all quasars must collectively share. Also notable is that, in effect, $D_0$ is akin to an experimental zeropoint, gauging a minimum correlation which still avoids QM-mimicry for antipodal, minimally acausal quasars: they cannot be {\it less} correlated than this.

\subsection{Simulated Sampling of Quasars}

A software to generate simulated observations of this quasar flux-correlation-distribution on the sky was created. It is written in Interactive Data Language (IDL), and called spooky\_correlation; available via GitHub \cite{Steinbring2026b}. It creates instances of correlation between pairs for observations taken at any random time, and then averaging over many trials. The results are illustrated in Figure~\ref{figure_probability}: 10000 realizations of $P$ for $M=0.87$ by randomly selecting residuals from a Gaussian distribution with $\bar{R}=0.67$ (grey points, left panel; distribution, right panel). Sometimes an excess correlation for a given quasar-separation is found, other times not. On average, it is apparent: the mean of trials at each angular separation indicated with a thick dark-grey curve, which (overall) differs from the expected mean of $\bar{P}=50\%$ by an increment just under 1.8\%, here subtracted - essentially incuring a small anisotropy. In this case, a fraction just over 4\% of sufficiently-correlated cases would rise above the critical, 75\%-detection level, indicated by those instances above that bar, within the peaks. Each is the probability that one random sample will be found there, although that predicts finding a pair. Put another way: if correct, in a large long-term random sample of causally-linked quasars uniformly distributed over the sky there might be 20 out of 10000 sensed with 10 biased samples each, and at least one such special pair is predicted for a separation within a few degrees of $65^{\circ}$.

Three key features of this potential correlation appear relevant to look for: First, there is a prominent enhanced-correlation peak near $65^{\circ}$ separation, seemingly sufficient to trigger a QM-mimicing result, as this is just over 25\% more likely than neutral. That would not be in conflict with previous experiments, which utilized quasars closer to $82^{\circ}$ separation \cite{Rauch2018}, where expected influence passes below the limit of corruption (less than 14\% biased). Second, despite preserving isotropy it is distinctly bimodal and asymmetric about $90^{\circ}$ in separation. That attribute is required to preserve an average neutrality of an ensemble across the whole sky, i.e. net correlation is counterbalanced at larger angles. And third: in this form, correlation is indeed suppressed and nearly vanishes at $180^{\circ}$, serving to confirm the applicability of antipodal acausal sources, as expected.

So far this is a geometric argument; apart from requiring isotropy while assuming correlated sources do exist, no conditions have been placed on the properties of those, such as flux or temporal behaviour on any timescale. There is no reason to expect that correlated fluctuations must happen on short timescales, instead possibly lasting more than a night or much longer. For example, if "unlucky" perhaps two may only happen to be in sync and visible within just one night every few years, so those are not likely to be discovered in practice. As stated, the proposed metric simply gauges how correlated any two quasars may be against the global average in their flux-differences, but now the defined passbands available in the study, and expected noise sources against such correlations with angular separation on the sky can be characterized, to verify a signal could be detected in a survey.

\subsection{Angular Dependence of Noise}

For simplicity, a binary condition of red/blue has been used to describe quasar correlation. But any such discrimination is really just a relative flux measurement over a suitably defined passband and time period: some greater/lesser number of photons are received, divided by integration time. And for this to be well defined, it is only necessary to consider if, during that time, the measured relative-flux of those photons between telescopes is dominated by local noise, to avoid spurious correlation. From the ground this could come from seeing, sky background and extinction, well known to depend on airmass, and so injecting some cosine dependence on errors with target-separation angle $\phi$, which must now be excluded against any inherent in the quasar fluxes.

\begin{figure}
\center{\includegraphics[width=7 cm]{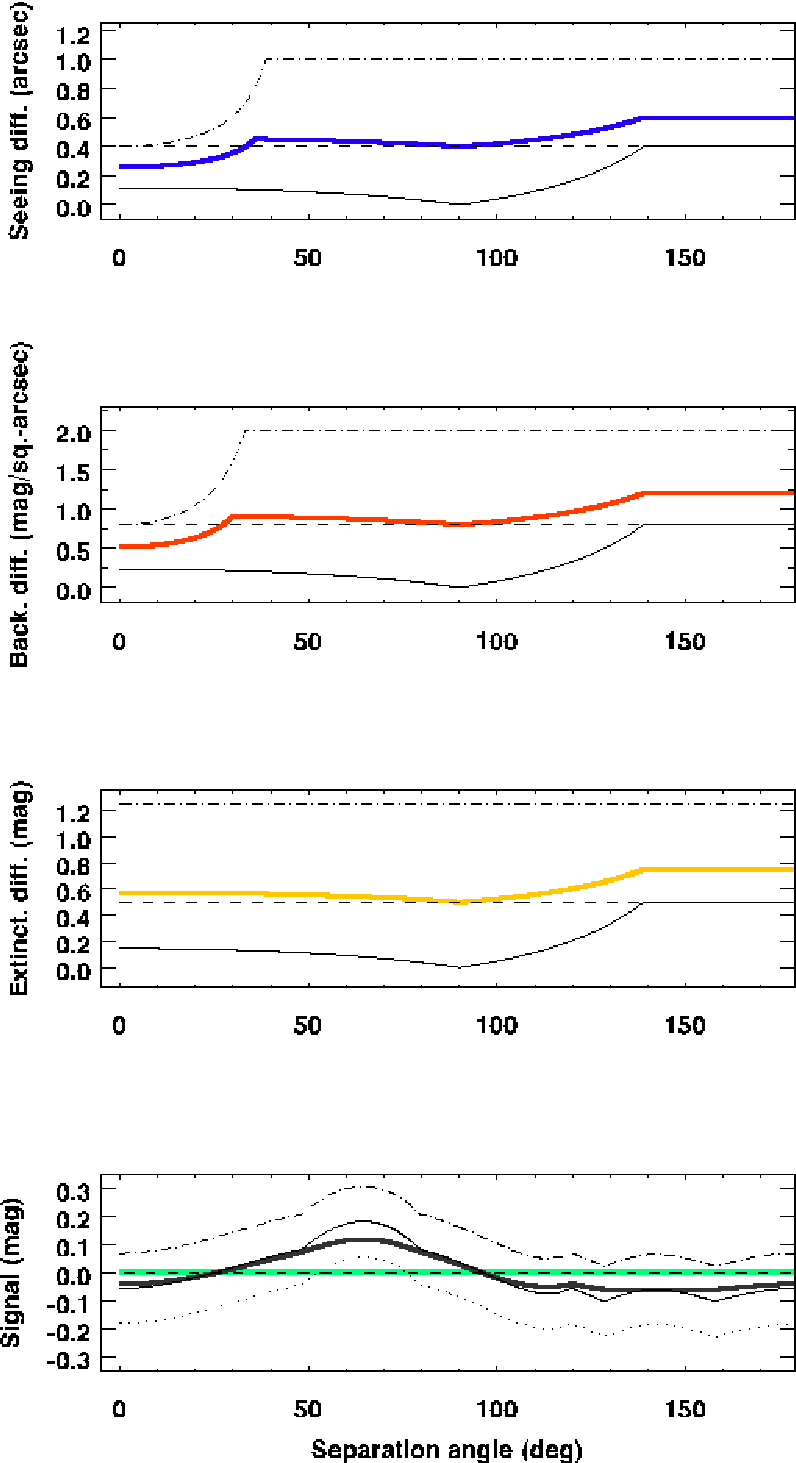}}
\caption{Model differences in seeing (blue), sky brightness (red), and extinction (yellow) between two positions on the sky up to $180^{\circ}$ apart, viewed from by Gemini. Lower limits (thin black curves) and upper limits (dot-dashed) are indicated, giving those means of coincident observations and long-term medians (dashed lines) for combining views from each site within a night, or otherwise. Below is the long-term difference of two quasars selected from a brightness distribution of width 1.0 mag, if their fluxes were correlated as per probability $P$ at $S/N=3$ (thin black curve; $5\times$: dotted, dot-dashed) or random residuals (thick, grey) against their averaged instrumental errors (thick, green).}\label{figure_models}
\end{figure}

Consider identical instruments at two sites with a geographic separation of $90^\circ$, that is, two sources viewed simultaneously at zenith are at right angles. (This is essentially the case for Gemini North and South: $95.5^\circ$ apart.) Under clear skies, atmospheric extinction increases linearly with airmass, inversely with zenith distance ($1/Z$), so to a target under an airmass of 2, the difference between it and another $\phi$ degrees away, is at its extreme $$E(\phi) = \Big{|} 2 - {\sqrt{2}\over{\cos{(\phi/2)}}} \Big{|},\eqno(9)$$or $$A = \alpha E,\eqno(10)$$where $\alpha$ is half the median extinction in magnitudes. A similar relationship can be found in sky brightness, $$B = \beta E,\eqno(11)$$and image quality $$C = \gamma E_\gamma,\eqno(12)$$where $\beta$ is in units of ${\rm mag}~{\rm arcsec}^{-2}$, and $\gamma$ is in arcseconds. The functional form of $E$ has a zenith distance more like a standard expression \cite{Kasten1965}, which is $Z=\cos{(\phi/2)} + 0.150 \times (93.885 - \phi/2)^{-1.253}$, and for seeing ($E=E_\gamma$) is perhaps better modeled with a weaker $Z^{0.6}$ power. And while one site is still enjoying better weather, the other might continue imaging only under 80\%-ile conditions, a relative factor $2.5\times$ poorer than the median. Those limits are shown in Figure~\ref{figure_models}, represented by a thin line and a dot-dashed line respectively. Later it will be shown that for Gemini $\alpha\approx 0.25~{\rm mag}$, $\beta\approx 0.40~{\rm mag}~{\rm arcsec}^{-2}$, and $\gamma\approx 0.38~{\rm arcsec}$; so those values are adopted in Figure~\ref{figure_models}. Their averages are shown as thick curves.

\subsection{Observationally Detectable Signal in a Survey}

Each observing night with Gemini likely includes both GMOS-N and GMOS-S imaging. So in long-term survey, many samples of a minute or so duration are taken near-nightly, all at random positions over the full sky. Although quasars are known to fluctuate on timescales of days to many years, and likely do so on timescales as short as minutes, as a group they have well-studied optical brightness distributions. So if those were sampled in a perfectly unbiased way from a distribution of width $\omega$ their maximal long-term difference as amplified by equation 2, normalized by peak $\hat{P}$ to its mean is $$S = \omega P (1-\hat{P})/\bar{P} \approx 0.27 \times \omega P,\eqno(13)$$in magnitudes. Notice how $S$ measures increased, correlated {\it differences} in fluxes against their mean in any sample of measurements for sources of some angular separation. In effect, a signal amounts to a higher dispersion relative to flatness with separation angle, and simply requires a sufficient number quasar samples to overcome noise per angular-separation bin, at minimum restricted by the instrumental error per sample.

The signal-to-noise ratio of detectable enhancement (flux-differences per angular resolution-element) over observational noise for $n$ samples thus has a form $$S/N \propto {\omega P \over{(\alpha E + \beta \gamma^2 E^3 + \epsilon)/\sqrt{n}} + \zeta},\eqno(14)$$relative to an ideal $\overline{S/N}=1/0.27$ limit, where $\epsilon$ is the photometric uncertainty and $\zeta$ is the bandpass zeropoint error, both in magnitudes. Note that, as they are maximal, $A+B+C$ are not added in quadrature, and fall off as the square-root of the number of samples, as does photometric uncertainty. Binning the data can enhance $S/N$ only until it reaches the zeropoint accuracy. For example, a 1-mag quasar-brightness distribution in the optical and zeropoints not typically better known than to 2\%, so the detectable effect in a randomly selected pair is limited to about $\omega / \zeta \lesssim 50$. Extended to a long-term average, the instrumental errors average out across the sky over many samples. Thus $E_\gamma=E=1$, which implies that (sampled over all angles) the effect slowly grows to $S/N\approx2.9$ again by $n=10000$ with photometric uncertainty fixed at $0.25~{\rm mag}$. 

This suggests that, despite being individually noisy, relative photometry of an unbiased quasar sample through many years could find the expected dependence with angular separation. For reference, 10 Gemini GMOS FoVs ($5~{\rm arcmin}$ across) spans roughly one degree. Averaged over the sky, a detection minimum of $S/N=3$ suggests a further constraint: the main peak and inflections in Figure~\ref{figure_probability} are about $30^\circ$ across (minima to minima), perhaps covered optimally by bins about $6^\circ$ wide, and the cusp at $\phi=120^{\circ}$ is minimally spanned by 3, and just sufficient to discern a difference between a peak at $64.6^{\circ}$ from $60^{\circ}$.

\section{Sample}\label{sample}

All publically-available GMOS-N and GMOS-S images were searched for instances of a known quasar falling into the field, starting from the beginning of regular operations through 2016, spanning 14.5 years. A difficulty with a direct search is that the file header information will include a target name selected by the observer, which may not necessarily correspond to any catalog. Also, this would exclude cases where an object happened to fall on the detector during the observation of another, defined target. And, instead the Million Quasars Catalog (MILLIQUAS), Version 4.8 of 22 June 2016 was cross-correlated with the full Canadian Astronomy Data Center (CADC) archive GMOS images. The MILLIQUAS is a compendium of published catalogs, primarily the Sloan Digital Sky Survey (SDSS) providing a redshift plus optical magnitudes (Blue: B, V or ${\rm g}$; and Red: ${\rm r}$ or ${\rm i}$) for each object \cite{Flesch2023}. No restriction on type was made, including gravitational lenses, but the likelihood of misidentifications is considered small.  There is also a small possibility of some of those being falsely-identified stars, effectively adding some negligible amount of noise. Even so, to ensure clarity, each identified quasar will be now be referred to as an "object."

The GMOS field of view (FoV) is approximately $5 \times 5~{\rm arcmin}^2$, with 0.075 arcsec pixels. Using the Common Archive Observation Model (CAOM) Table Access Protocol (TAP) web service and custom Python scripts, a search inside a 5 arcmin radius for each object was conducted, which produced 28374 cases where an $0.1 < z < 6$ object was within a GMOS field in ${\rm g}$, ${\rm r}$, ${\rm i}$, or ${\rm z}$ (similar to the SDSS filters) through the full range of right ascension, and between $-79^{\circ}$ and $+82^{\circ}$ declination; deleting those corrupted or otherwise unusable yielding 20514 objects (${\rm g}$: 4691, ${\rm r}$: 6709, ${\rm i}$: 6776, ${\rm z}$: 2338), and a typical exposure time of 150 s. Gemini grades each science frame by clear-sky-fraction bins from CC20 (best 20\%-ile) to CCAny (all conditions). This represents 11483 unique (although potentially repeated) objects having a mean redshift of 1.48, no noticeable dependence on sky position, and obtained under all conditions under which the telescopes were operational.

Finally, a selection was made to ensure good data quality in each observation, following proceedures similar to another GMOS long-term survey \cite{Steinbring2022}. Every object frame was searched for a comparison star from the Fourth U.S. Naval Observatory CCD Astrograph Survey (UCAC4) to serve as a photometric calibration, with published SDSS ${\rm r}$ magnitude. There were 16774 frames that had such a suitable, unsaturated star. This also provided each an image quality criterion; image Point-Spread Function (PSF) Full-Width-at-Half Maximum (FWHM) was extracted for each star. The FWHM of the object (also confirmed to be unsaturated) was checked against that; in just a few cases where it occurred, the smaller of the two was taken.  A lower FWHM limit of 0.20 arcsec was applied to ensure that only real images were obtained, without artifacts, and so it reflects seeing conditions. After this, a total of 9317 good samples remained after culling to cloud-cover conditions better than CC80, or usable, with less than 2 mag of extinction.

\section{Photometry}\label{photometry}

Synthetic aperture photometry was carried out on the full sample (objects and comparison stars) using a 4 arcsec diameter aperture throughout. Roughly 20 ${\rm arcsec}^2$ postage stamps sections were downloaded from the database. The positional accuracy of the frames was found by inspection to not always be better than 2 arcseconds. During the FWHM measuring step, a centroiding algorithm located the central pixel position and then sub-pixel shifted the aperture prior to obtaining the flux. Median sky backgrounds for each frame were subtracted after applying the appropriate detector gains and filter zeropoints, as published on the Gemini webpages. The detectors, and their layout in the focal plane changed at certain times, either 4 or 6 chips across the FoV with separate amplifiers and gaps between them, and each detector has a slightly different amplifier response. On average these are $28.11$, $28.31$, $28.16$ and $27.17$ mag in ${\rm g}$, ${\rm r}$, ${\rm i}$ and ${\rm z}$, which fluctuated over that time with deviations of 0.12, 0.02, 0.12 and 0.22 mag respectively.

\begin{figure}
\center{\includegraphics[width=7 cm]{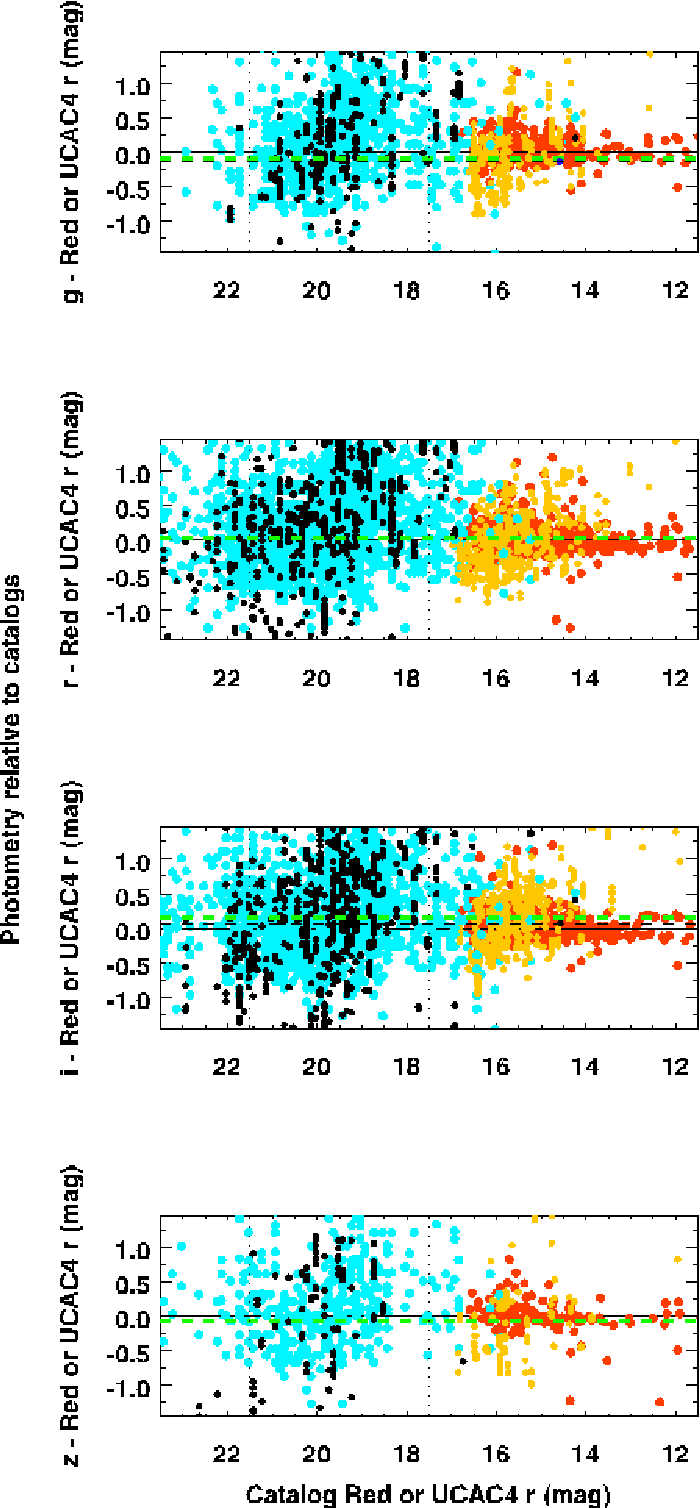}}
\caption{Differences between observed and catalog magnitudes for objects (black) and comparisons (yellow) in the four filters studied. The relative colours of those against MILLIQUAS Blue (light blue) and UCAC4 ${\rm i}$ (red) are also shown; comparison colours (thick-dashed green) are neutral, as are the objects (dashed black). Those last have been limited to magnitude cutoffs (vertical dotted lines), serving to avoid a sample colour bias; if perfectly Gaussian each would be zero.}\label{figure_by_filter}
\end{figure}

\begin{figure}
\center{\includegraphics[width=7 cm]{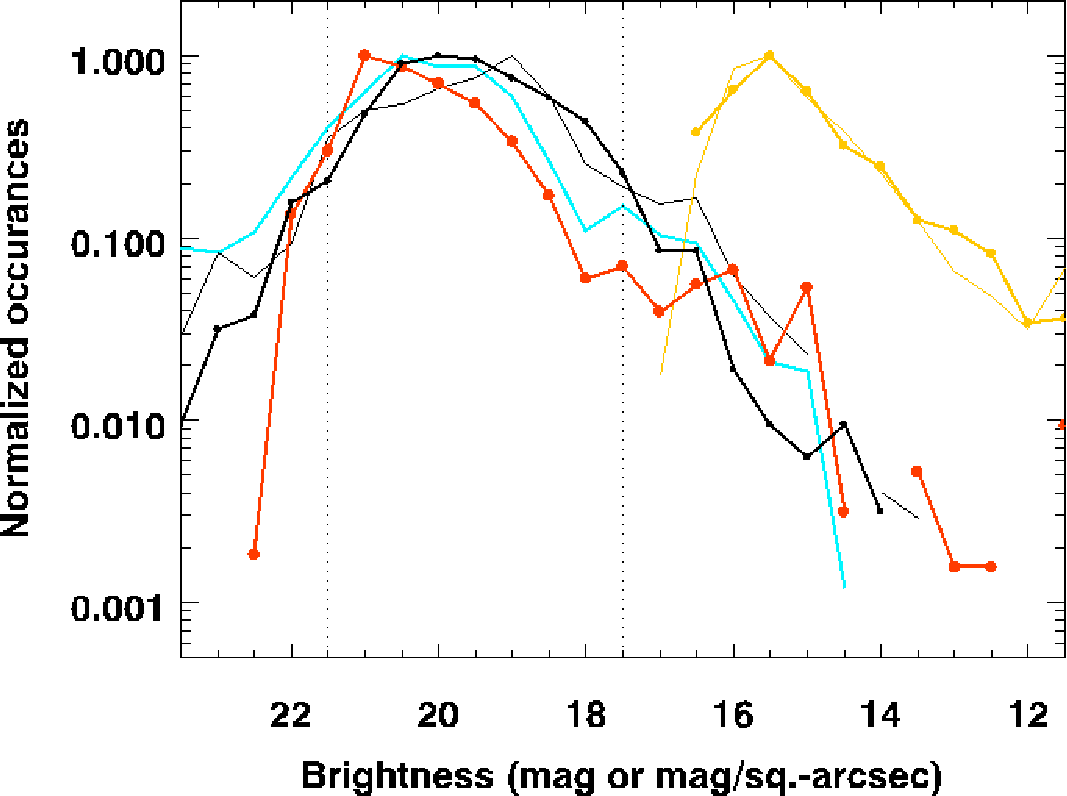}}
\caption{Histograms of sky background (red), object (black), and comparison star (yellow) magnitudes, normalized to peak occurrences. Observed values are comparable to catalog MILLIQUAS Blue (light blue) and Red (thin black) vas are comparisons to catalog UCAC4 r (thin yellow).}\label{figure_magnitude_histogram}
\end{figure}

All resultant photometry was corrected for atmospheric extinction using the calibration stars, where each of those was relative to its UCAC4 ${\rm r}$ brightness and a mean filter correction; results are shown in Figure~\ref{figure_by_filter}. This was calculated using the 237 observations with complete photometry in MILLIQUAS catalogued B, V, ${\rm r}$ and ${\rm i}$ magnitudes, which allows the calculation of a mean sample colour shift to B, V, ${\rm r}$, and ${\rm i}$ of 0.84, 0.49, 0.34, and 0.19 mag: g was interpolated as 0.66 mag and ${\rm z}$ extrapolated to 0.18 mag. Of these, 124 were cases of objects included in the UCAC4 catalog, all brighter than 17.5 mag. Correcting stellar colours so that the average sky background difference across the full sample is zero in ${\rm r}$ yields corrections of 0.50, 0.00, -0.27, and -0.22 mag in ${\rm g}$, ${\rm r}$, ${\rm i}$, and ${\rm z}$ filters respectively; although less of a concern in ${\rm r}$ and ${\rm i}$ data, the mean values of the resultant sample in all four filters are very close to neutral. The objects were also shifted by the same mean filter correction to ${\rm r}$ and uniformly taken as a differential from their catalog magnitudes. In this way, all photometry is relative to the sample average of ${\rm r}=20.0$ mag; photometry is sky-background-limited, with some down to the expected $5\sigma$ point-source limit of 23 mag: see Figure~\ref{figure_magnitude_histogram}. A lower object cutoff at 21.5 mag excludes those fainter than the mean sky surface brightness, avoiding a colour bias.

\begin{figure}
\center{\includegraphics[width=6 cm]{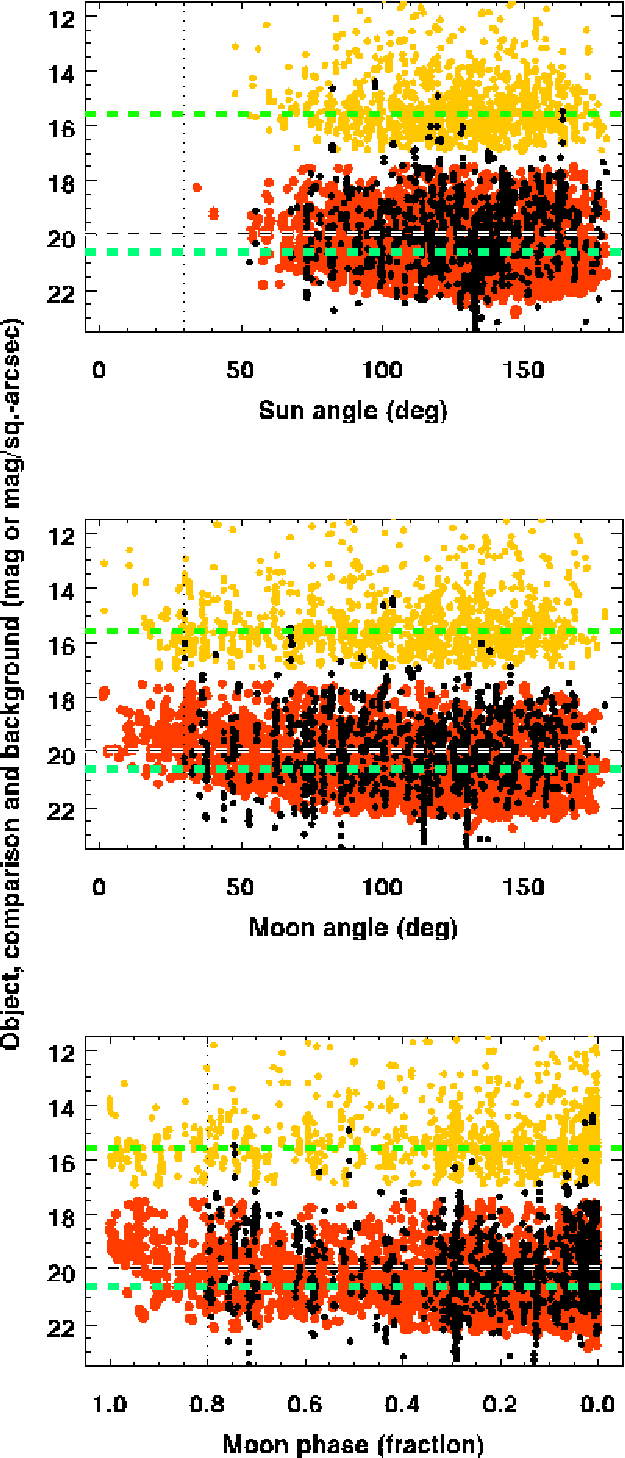}
\hspace{1 cm}\includegraphics[width=8 cm]{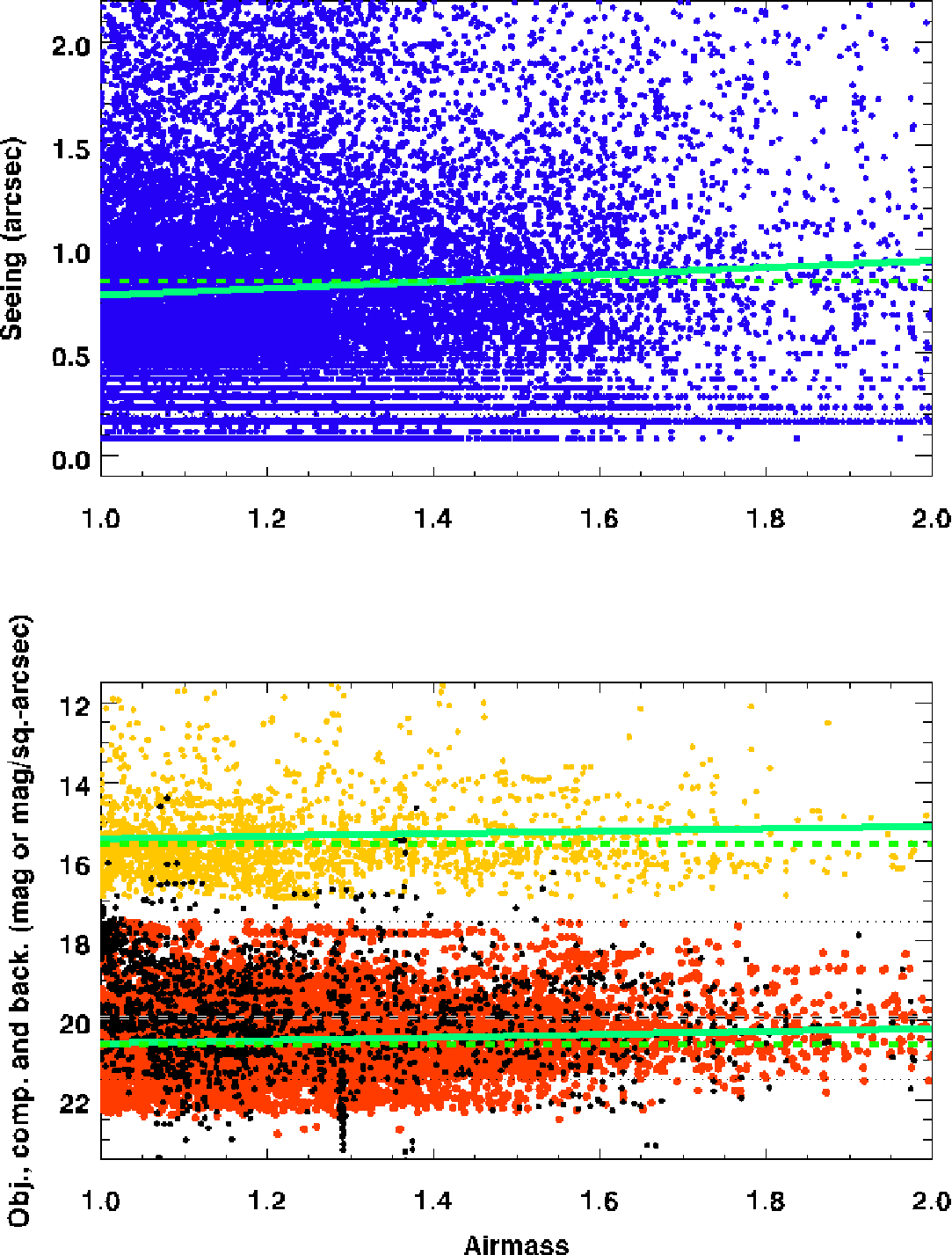}}
\caption{Left column of panels from top: object, comparison and sky background with angle to Sun and Moon, and Moon phase; a proximity limit of $30^\circ$ and 80\% full Moon (horizontal dotted lines) ensures skies were suitably dark; medians shown as dashed horizontal lines, against comparisons and sky background (thick, green). Right panels: All object magnitudes obtained (filled black circles) by airmass, with the UCAC4 ${\rm r}$-band magnitudes for comparison stars in each field (yellow) and their background sky brightnesses (red); sample limits have been applied (horizontal dotted lines). Above: image quality estimates for each frame (blue); data are spurious below a cutoff at 0.20 arcsec (horizontal dotted line). Medians and linear-least-square fits are shown for comparison (thick, green).}\label{figure_object_by_airmass_and_with_sun_and_moon}
\end{figure}

The weak influence from sky brightness is illustrated in Figure~\ref{figure_object_by_airmass_and_with_sun_and_moon}. Only mild angle and sky-brightness restrictions were employed, with a uniform upper cutoff of $17.5~{\rm mag}~{\rm arcsec}^{-2}$. To meet it, objects must have been at least $30^\circ$ from the Sun or Moon, during phases less than 80\% full. This agrees well with expectations from the linear model. Linear least-squares fits to extinction (comparison magnitudes), sky brightness and seeing (image quality) are shown in thick green (medians: dashed), giving $\alpha=0.25$ mag, $\beta=0.40 ~{\rm mag}~{\rm arcsec}^{-2}$, and $\gamma=0.38$ arcsec.

\section{Analysis and Results}\label{results}

\begin{figure}
\hspace{-2cm}
\includegraphics[width=19 cm]{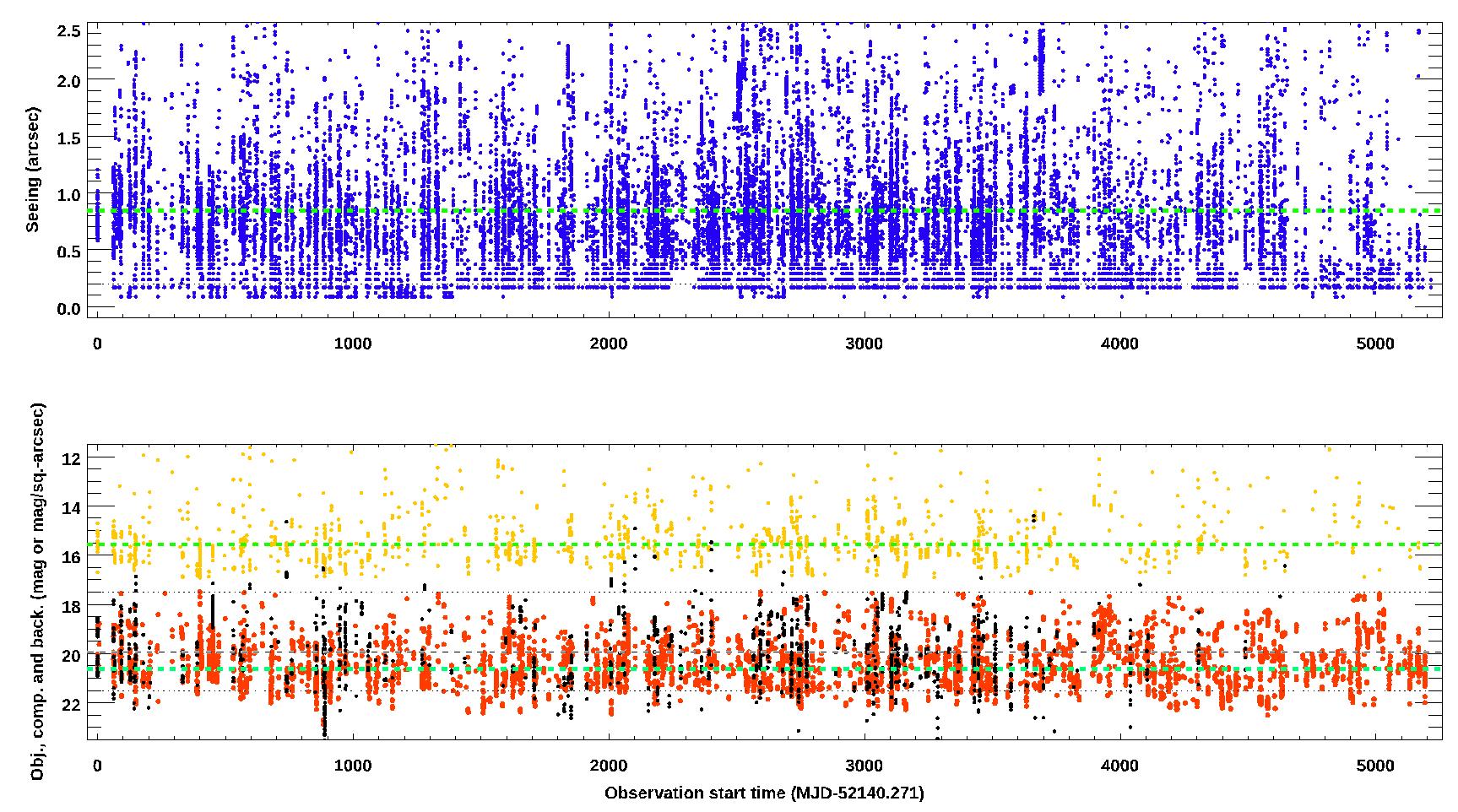}
\caption{All output data plotted in time: objects (black), comparison stars (yellow) and background (red) in magnitudes (bottom panel) and seeing in arcseconds (blue; top panel), consisting of 11483 objects sampled over 5200 nights; effectively unbiased within each of those for all separations from zero to $180^{\circ}$, and from about one minute up to 17-hour timescales.}\label{figure_object_comparison_background_and_seeing}
\end{figure}

Observations were recorded with their object sky position, UTC exposure start time, and duration from each frame header. An IDL code was written to perform the following cross-comparison prescription: for each observation, all earlier ones were searched to find if the exposure duration overlapped, calculate the object-to-object separation angle, and take the relative differences in magnitudes, final minus initial (although reversing order is equivalent, relative to their mean). This last output is the signal to be reported. Perfect overlap, or perfect coincidence, was assigned a fraction of unity, and all less - down to zero - by the fraction of duration overlap. Negative values then spanned back in time to include all earlier exposures in a widening delay window. For example, an initial check was to find if another observation occurred during the same observational "day" at the combined Gemini telescopes, which varies during the year, but is about 11 hours plus the timezone difference, or 17 hours. The largest possible delay is back 14.5 years, to the beginning of records; providing $2\times10^6$ total sample-pairings. The full dataset is plotted in Figure~\ref{figure_object_comparison_background_and_seeing}.

\begin{figure}
\center{\includegraphics[width=9 cm]{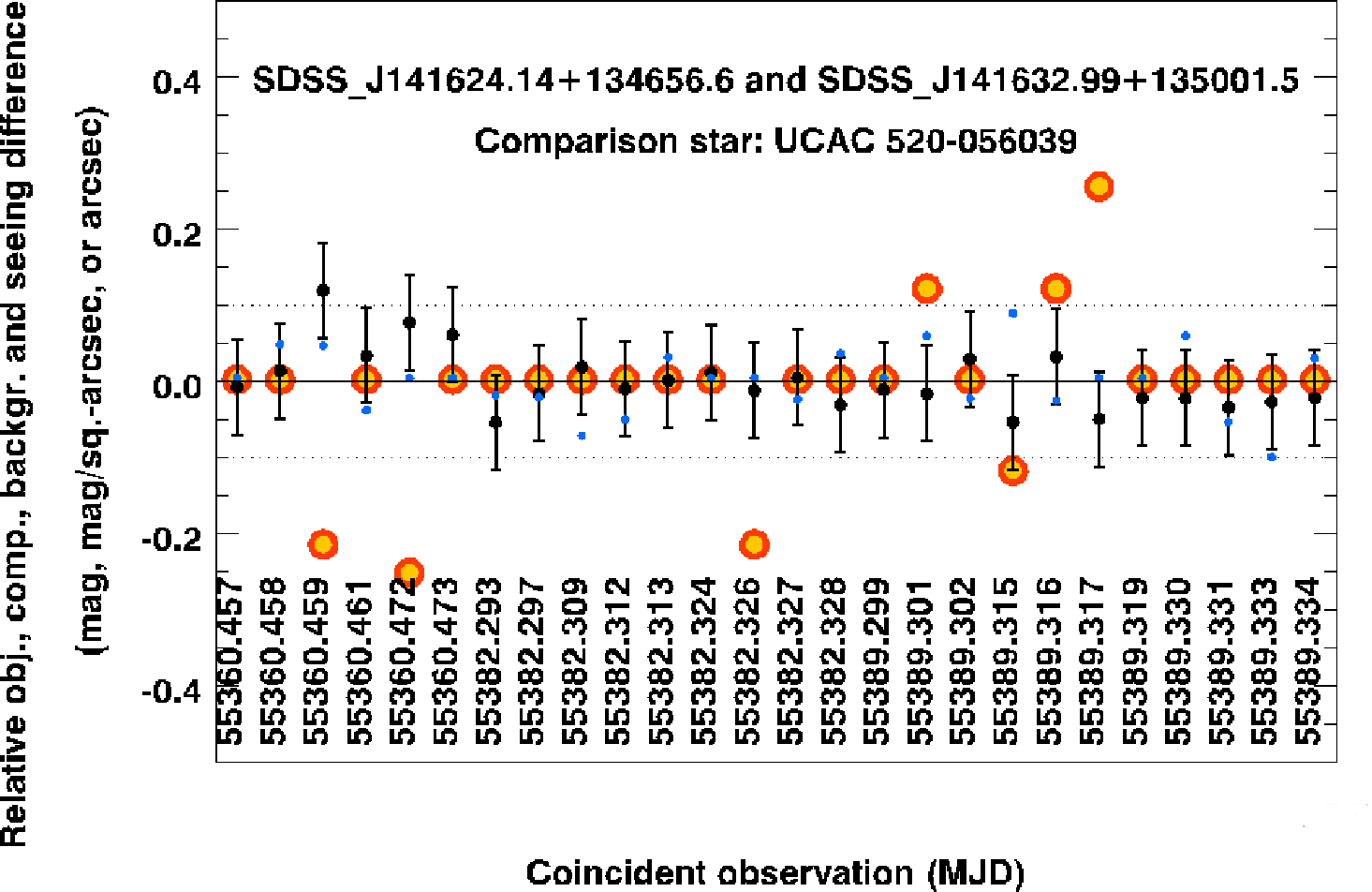}}
\caption{Differences in object, seeing, sky background and extinction for two particular quasars falling inside the same frame, spanning one month. These coincident observations demonstrate the good data quality and photometric accuracy obtained for the sample.}\label{figure_differences_special}
\end{figure}

Some fully coincident observations did occur, that is, object pairs with fractional overlap of unity (310 times). One such case with multiple repeated samples is shown in Figure~\ref{figure_differences_special}, for quasars SDSS J141624.14+134656.6 and SDSS J141632.99+135001.5, observed with GMOS-North 26 times together over the course of a month (the MJD observation times are indicated), when those two happened to fall inside the same frame. In most cases, the comparison star was the same (UCAC4 520-056039) and so there is no difference in measured sky background (red circle), extinction (yellow) or seeing (blue dots). Those differ, however, in the few cases where a different calibration star was found (automatically) for the other. Spanning these observations, the uniformity of the photometry is remarkably stable. It is evident that photometry of quasars near the mean sample brightness (catalog magnitudes of 19.7 mag and 19.8 mag) was carried out over multiple observations over month-long timescales accurate to 0.10 mag, which is considered the limit of significant measurable differences for the rest of the sample.

Several repeat observations having brightness differences larger than the significance limit are provided in Table~\ref{table_special}, ordered by the span of samples. Only those with at least 10 occurrences are retained; uncertainties reported are standard deviations. Each object-to-catalog difference reported here is the average of those samples, not the signals relative to all other individual samples. These were observed from both North and South, and through the full range of catalog brightness. There are some notable cases of bright (up to 17.5 mag) quasars, although none is reported in the literature as a known variable. For example, 3C 186.0 ($z=1.07$) is a well-studied radio-loud source with a prominent jet \cite{Chiaberge2017}: observed 11 times beyond the significance limit, on average $0.40$ mag less than its catalog Red magnitude (and even brighter than its SDSS r value of 17.88). The GMOS filter-bandpasses themselves did not change, but emission lines falling in or out of those sampled could affect that. The largest discrepancies from the catalog are over a magnitude, occurring in fainter objects, as expected.  Monitoring of targets might reveal, e.g. intrinsic, intra-day brightness fluctuations of emission lines. 

\begin{table}[t]
\hspace{-1.75cm}
\begin{tabular}{lrrlccrccr}
\hline
\hline
                             &R.A.            &Dec.            &            &Cat.       &Obj.               &Diff.      &         &Start           &Span\\
Name                         &(deg)           &(deg)           &$z$         &(mag)      &(mag)              &(mag)      &$n$      &(MJD)           &(days)\\
\hline
SDSS J090752.02+135829.2     &136.96675       & 13.97480       &1.7         &20.2       &$19.19\pm0.41$     &-1.01      &12       &56040.040       &   0.0239\\
SDSS J081839.27+574750.6     &124.66362       & 57.79739       &1.632       &20.6       &$19.85\pm0.12$     &-0.75      &12       &55153.474       &   0.0241\\
SDSS J081838.61+580235.7     &124.66088       & 58.04328       &1.2         &21.0       &$20.64\pm0.07$     &-0.36      &12       &55158.523       &   0.0248\\
SDSS J120835.93+020559.4     &182.14975       &  2.09986       &2.302       &20.2       &$20.35\pm0.06$     & 0.15      &10       &55270.223       &   0.0267\\
SDSS J120843.80+020840.8     & 82.18250       &  2.14467       &1.0         &19.6       &$20.46\pm0.06$     & 0.86      &10       &55270.223       &   0.0267\\
SDSS J111010.37+011423.8     &167.54321       &  1.23994       &0.4         &22.0       &$20.85\pm0.72$     &-1.15      &14       &55587.294       &   0.0460\\
SDSS J002251.40+155652.3     &  5.71417       & 15.94789       &2.235       &20.4       &$19.85\pm0.17$     &-0.55      &17       &55774.540       &   0.0702\\
3C 186.0                     &116.07279       & 37.88811       &1.069       &17.5       &$17.10\pm0.11$     &-0.40      &11       &54151.217       &   0.1140\\
SDSS J042619.41+165726.8     & 66.58092       & 16.95747       &0.5         &20.4       &$18.65\pm0.13$     &-1.75      &30       &52591.485       &   1.0454\\
2QZ J112636.9+003454         &171.65408       &  0.58208       &0.550       &17.9       &$18.33\pm0.30$     & 0.43      &10       &54918.397       &   1.0872\\
MC 1043-291*                  &161.41917       &-29.45722       &2.128       &18.9       &$18.45\pm0.21$     &-0.45      &18       &53795.325       &   1.8113\\
NBCK J140849.77-010850.5*     &212.20737       & -1.14739       &2.0         &21.4       &$20.80\pm0.40$     &-0.59      &29       &53848.160       &   2.0415\\
SDSS J135335.92+401723.1     &208.39971       & 40.28975       &1.9         &20.9       &$20.05\pm0.16$     &-0.85      &23       &52701.576       &   2.0511\\
SDSS J003027.98+261804.2     &  7.61658       & 26.30119       &1.534       &20.1       &$19.88\pm0.05$     &-0.22      &10       &55471.265       &   4.1237\\
VA-562                       &  7.60862       & 26.28028       &0.269       &18.2       &$18.92\pm0.05$     & 0.72      &22       &55471.263       &   4.1261\\
SDSS J095155.67+220947.5     &147.98200       & 22.16319       &0.634       &17.9       &$17.67\pm0.12$     &-0.23      &10       &55296.298       &   9.9686\\
SDSS J095205.98+221018.8     &148.02492       & 22.17192       &2.627       &20.3       &$19.99\pm0.12$     &-0.31      &10       &55296.298       &   9.9686\\
SDSS J141624.14+134656.6     &214.10063       & 13.78242       &2.259       &19.8       &$20.88\pm0.07$     & 1.08      &27       &55360.457       &  28.8773\\
SDSS J141632.99+135001.5     &214.13750       & 13.83375       &1.0         &19.7       &$19.93\pm0.09$     & 0.22      &29       &55360.457       &  28.8773\\
SDSS J023639.93+282308.2     & 39.16642       & 28.38561       &1.9         &21.1       &$20.12\pm0.07$     &-0.98      &12       &54707.543       &  35.9549\\
SDSS J023653.25+282142.3     & 39.22188       & 28.36178       &1.0         &19.9       &$19.24\pm0.06$     &-0.66      &12       &54707.543       &  35.9549\\
IXO 10                       & 50.66833       &-37.27778       &0.515       &19.5       &$20.04\pm0.36$     & 0.54      &15       &54760.314       & 308.0033\\
IXO 69                       &190.90213       & 11.50256       &1.195       &18.9       &$18.18\pm0.19$     &-0.72      &15       &54180.450       & 682.9031\\
LBQS 1308-0104               &197.83021       & -1.34192       &2.620       &17.5       &$18.18\pm0.20$     & 0.68      &11       &52267.626       & 730.0214\\
SDSS J002235.96+001850.0     &  5.64983       &  0.31390       &1.6         &20.1       &$20.62\pm0.27$     & 0.52      &23       &52141.458       &1383.9300\\
CXOMP J01527-1359            & 28.18250       &-13.98361       &0.821       &20.9       &$20.46\pm0.35$     &-0.44      &10       &52474.553       &2244.6849\\
\hline
\hline
\end{tabular}
\caption{Multiple Significant Object Differences Ordered by Span of Samples\label{table_special}}
\end{table}

\begin{figure}
\center{\includegraphics[width=8 cm]{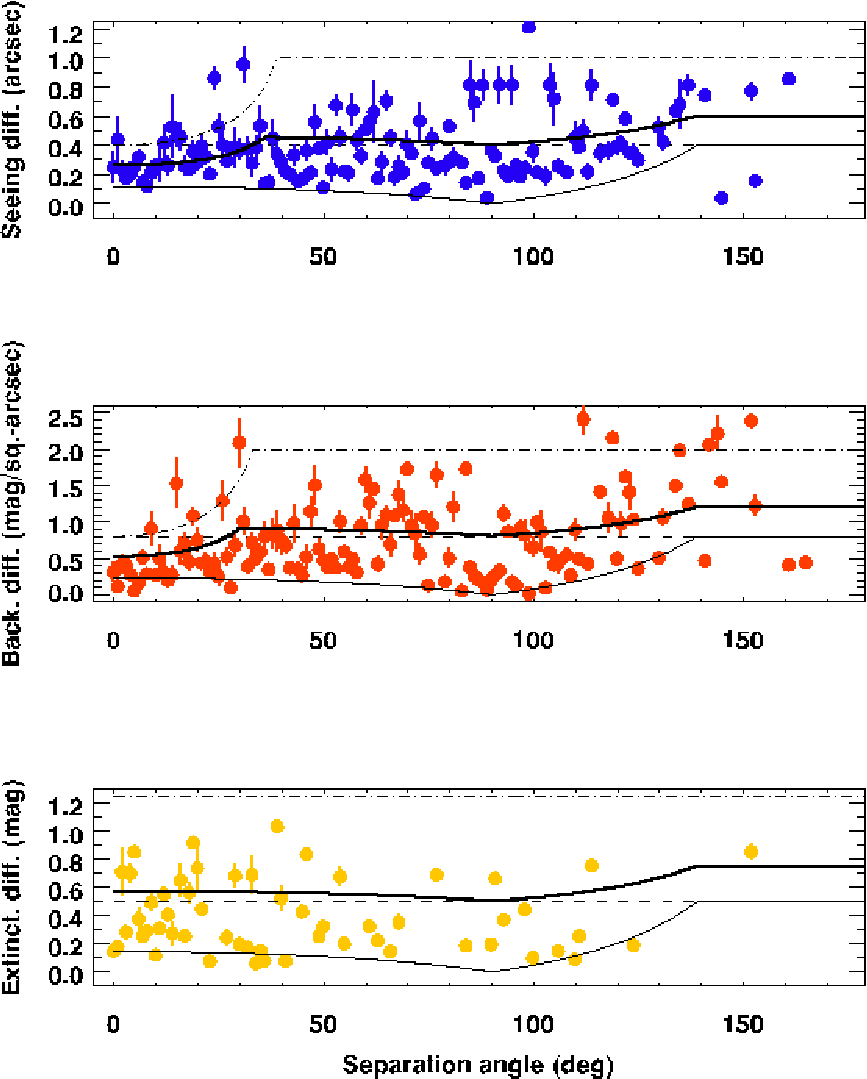}}
\caption{Differences in seeing, sky background and extinction occurring over a single day; a nightly window starting at twilight at Cerro Pachon and lasting approximately 17 hours. Data are averaged in 1-degree-wide bins, with $1\sigma$ error bars.}\label{figure_differences_one_day}
\end{figure}

As predicted, there are pairs separated by about $60^\circ$ found to be significantly different ($\geq 0.1$ mag) from their relative cataloged-differences. For example, MC 1043-291 and NBCK J140849.77-010850.5 ($z\geq 2$; asterix by names in Table~\ref{table_special}) are discovered in multiple samples spanning days (although not both taken together over the same nights). The samples of all other similarly-separated pairs significantly differing occured less frequently (than 10 times), and were taken in different nights. There was no case of sampled differences {\it between} the two telescopes captured during a single-day window, i.e. simultaneously at large angles. Even so, those nights with just one object in the GMOS FoV (either North or South; not both), are shown in Figure~\ref{figure_differences_one_day}. There are some outliers, for instance, occasions of particularly good seeing at low elevation. But it can be seen how expectations of intra-day variation in seeing, sky background, and extinction values - when a comparison star suffered less than 2 mag of cloud - are consistent with observations, giving confidence that samples separated by many nights and up to $180^{\circ}$ can be safely intercompared, without an angular bias.

\begin{figure}
\center{\includegraphics[width=14 cm]{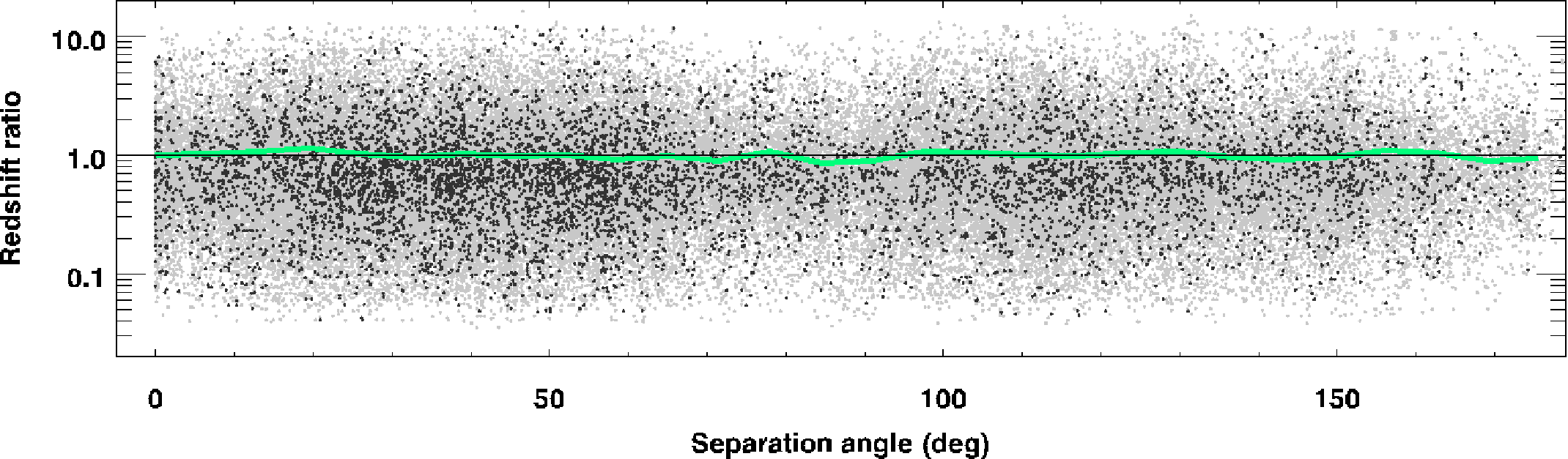}}
\caption{Ratios of redshifts between objects ordered by separation angle, for all cross-correlated observations (grey points) and those above significance (black); averages in six-degree-wide bins (thick, green), close to an expected perfect median of unity.}\label{figure_fraction_with_angle}
\end{figure}

Full cross-correlation of the entire sampled redshift catalog also indicates no evident bias in sample selection. Figure~\ref{figure_fraction_with_angle} shows the ratio of redshifts of any two objects as a function of angular separation. Although the visibility of objects from the two sites necessarily results in density variation of this distribution, it is fair, with no average redshift difference across the sample (thick, green). There were just 274 objects with $z\geq 3.65$, distributed over the whole sky, which was an insufficient sample to provide a meaningful comparison of just those; none were of truly acausal pairs separated by $180^{\circ}$.

\newpage

\begin{figure}
\center{\includegraphics[width=9 cm]{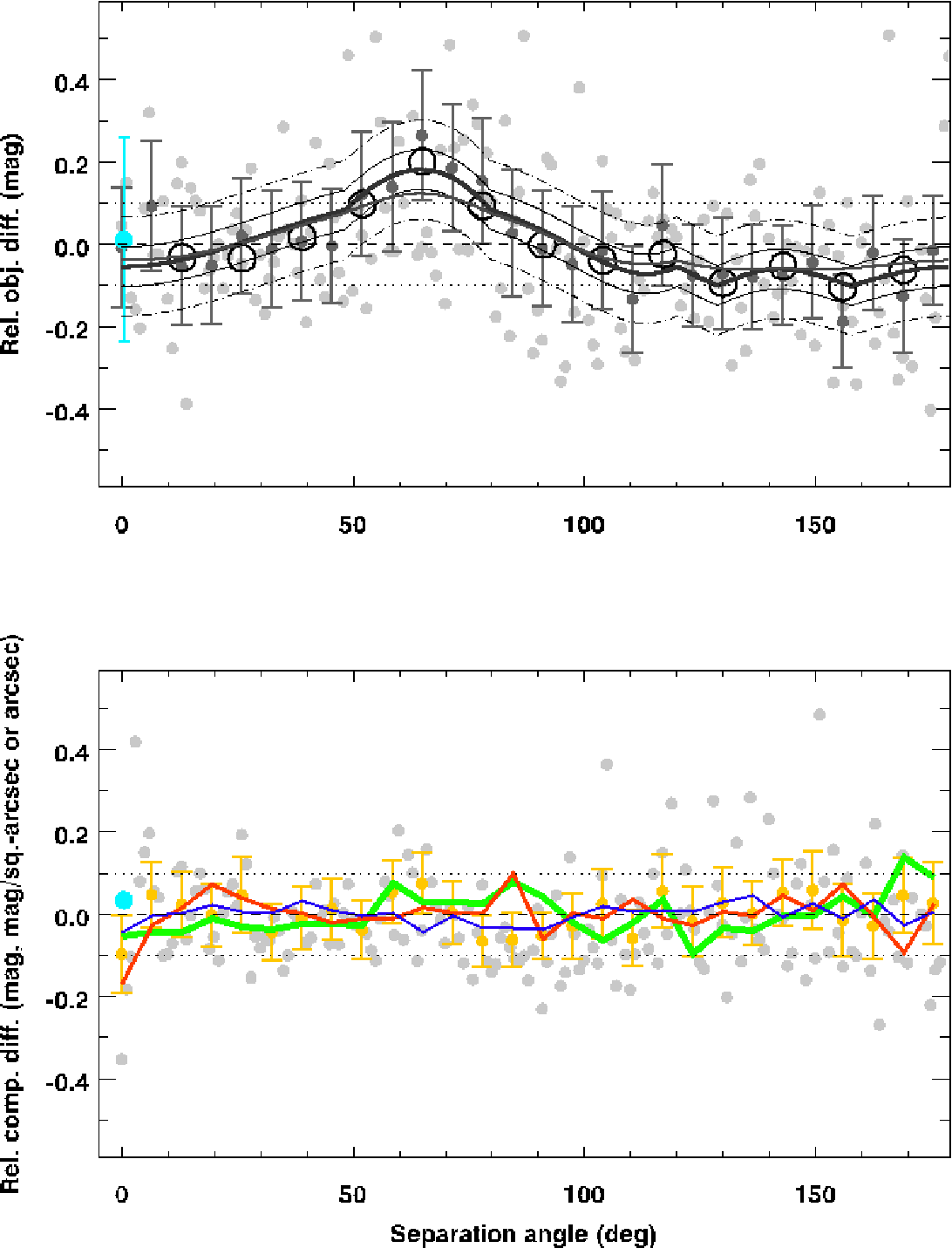}}
\caption{Object (light grey; top panel) and comparison-star differences (light grey: bottom panel) with angular separation in degrees averaged in 6-degree-wide bins (dark grey) within 68\%-confidence error bars, relative to the overall mean (dashed). Light-blue filled circles indicate the coincident cases only. Bottom panel: binned results for comparison stars (yellow, with error bars); the standard deviation of cross-correlated differences is 0.024 mag. Compared to the significance limit (dotted) the differences in catalog magnitudes are flatter (thick, green) as are sky brightness (red) and image quality (thin, blue). Top panel: signal is consistent with correlation as predicted by equation 13 (thick black curve) and open black circles show averages over two optimal angular bins, falling within twice the zeropoint error of that (thin black curves) and so the sensitivity limits of the photometry; $5\times$ the threshold-limits for residual correlation (dot-dashed) are also shown.}\label{figure_with_angle_final}
\end{figure}

The final results are shown in the top panel of Figure~\ref{figure_with_angle_final}: the most interesting being that for the full sample there do seem to be significantly-correlated object differences with angular separation. Only cases where objects were within 2 mag of each other and having a relative Blue-Red colour difference less than 0.20 mag were retained. These are displayed as the relative differences in the objects, averaged in 1-degree-wide bins (grey) and 6-degree-wide bins (black). The error bars are their $1\sigma$ standard deviations. The results of just the coincident pairs are shown in light blue, that is, those cases where two objects fell inside the FoV. Displayed in the bottom panel are the differences of comparison stars, for reference; seeing (thin, blue) sky brightness (red), and the catalog magnitudes (thick, green) are overplotted. These are all relatively flat (the standard deviation in comparison magnitude differences is 0.024 mag), which makes the comparison to equation 13 (top panel: thick black curve) remarkable, in agreement at all optimally-binned angles and strongly against flatness (dashed), for the deviation expected, at $S/N=3$. It should be emphasized that this is not a fit, and all parameters were specified in Section~\ref{introduction} by the sample conditions. But analysis of the cumulative distribution functions do confirm a good match: the Anderson-Darling (AD) test statistic is under 0.004, and it rejects the null-hypothesis of flatness at 97.7\% probability (${\rm AD}=0.886$, p-value of 0.023). Results are similar, with expectedly more scatter, if only ${\rm r}$-band frames are used.

\begin{figure}
\center{\includegraphics[width=11 cm]{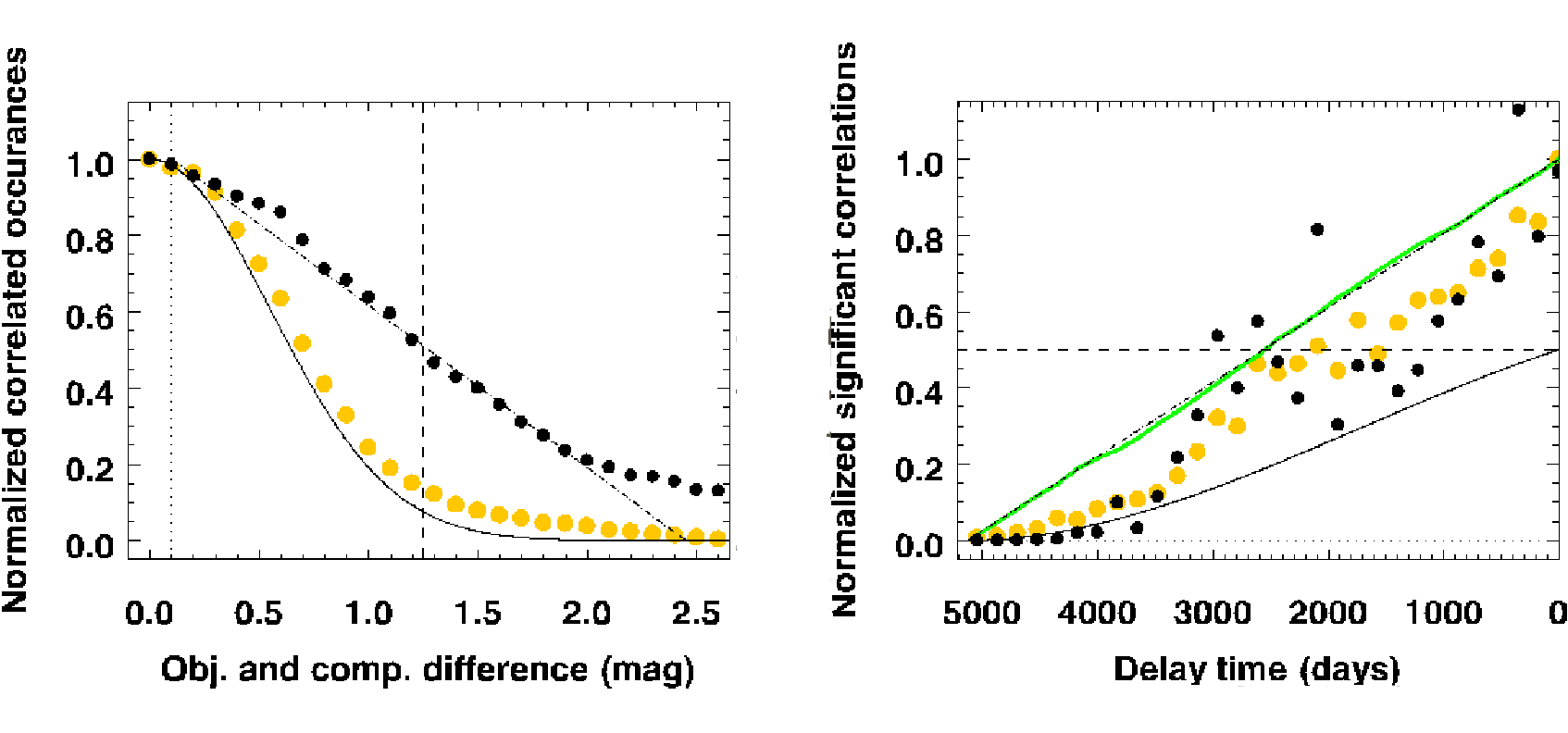}}
\vspace{-0.25 cm}
\caption{Left panel: normalized occurrences of differences between objects (black) and comparisons (yellow).  Right panel: same for significant differences only, as a function of delay time between correlated occurrences in days; thick green curve is the result when the order of data is randomized.}\label{figure_significant}
\end{figure}

These deviations of differences in object brightnesses are significant relative to the measured errors; they cannot be accounted for by photometric uncertainty. A significance limit of 0.10 mag is indicated by a horizontal dotted line in Figure~\ref{figure_with_angle_final}. The enhancement as per equation 13 near $65^{\circ}$ is secure; structure beyond $120^{\circ}$ perhaps less so, but still consistent with the data. The distribution of these differences is displayed as well in Figure~\ref{figure_significant}. A vertical dotted line (left panel) shows the limit of 0.10 mag; a thin black curve is a Gaussian of width 0.25 mag, consistent with purely photometric error; a dashed line is the expected difference for a uniform sample of width 1.25 mag. This is intuitively the limit one would expect, if the sample was randomly drawn from the same distribution, 1.00 mag wide.  Note the four or five instances near differences of about 0.50 mag that are above this line, so occurring slightly more often than one would expect from a randomly drawn sample.

With reversed seasons between North and South hemispheres, one would expect some seasonal dependence on when objects were observed, and so although they may be selected on the chance they appear in either telescope FoV this selection will not be uniform over the year. The right-hand panel of Figure~\ref{figure_significant} plots averages in 6-month-long bins for those cases which were beyond the significance limit from the sample (black). The thin black curve shows the result if half the sample were selected, decaying geometrically back towards the beginning of the window. The yellow circles are the comparisons for the full sample, the thick green line is the same except sample times have been randomized across the entire 14.5 years. Note this resampling has no effect on the results displayed in Figure~\ref{figure_with_angle_final}; those are absolute values, so the order in which the differences were taken is not relevant, and those have already been cross correlated for the full sample. It is interesting that there appear to be periods when it was more probable than random that there would be significant differences between two objects. The long timescales of those suggests that sensing correlation requires many repeat samples spanning years.

\section{Summary and Future Directions}\label{summary}

Guided by the geometric condition that a QM experiment employing antipodal quasars can provide acausal, and presumably uncorrelated photons, a gedankenexperiment where that is carried out simultaneously for every two quasars in the Universe has been considered. Applying the holographic principle, and demanding isotropy, this yields an expected correlation at lesser angular separations: in effect, a quasar found to brighten presages another found dimmer. Some residual flux (anti-)correlation should be detectable for unbiased random sampling across the sky, if not photon-by-photon, at least on the timescale of minutes to hours, or longer. By mining the full GMOS-North and GMOS-South archive, approximately 30000 broadband ${\rm g}$, ${\rm r}$, ${\rm i}$ and ${\rm z}$ quasar images were found, with many repeated; the vast majority of which are merely serendipitous: targeting an unrelated object in the field.  Photometry of each frame, which includes a stellar calibration for uniform data-quality selection, allows a careful analysis of the impact of quasar colour, variable sky conditions and airmass (sky background, extinction, and image quality) without explicit target or observer bias. Almost 10000 sufficiently deep observations for Gemini GMOS North and South, complete with a nearby unsaturated UCAC4 star, allowed photometry with a global zeropoint uncertainty of about 2\% over a sample spanning 14.5 years. These comprise roughly 2 million observational pairings, which in their aggregate have 0.25-mag $1\sigma$ uncertainty within 6-degree-wide bins. This is sufficient to show an expected lack of flatness in relative object flux-differences - an added dispersion in brightness - with angular separation. That residual connection between distant sources is less than could cause a conflict with causality, nor does it obviate previous QM tests. Still, it is a surprising result: distant quasars seem to spookily "know" about each others' fluxes. 

The mean redshift of the sample is 1.48, so if real, such a connection would have implications for AGN-formation histories, and more broadly for their galaxy hosts. Those objects are now cosmologically distant from each other, and so any coordination that could cause it happened at an earlier epoch. And because the expected correlation was arrived at by application of the holographic principle together with Bell's theorem, this hints at a potential, even deeper physical tie: perhaps quantum gravity on a cosmic scale may actually underlie a correlation between quasar fluxes. But it also has a practical consequence for applying QM experimental tests here on Earth. This dataset provides only a small number of truly coincident observations (yielding just those cases where two quasars were in the same GMOS FoV) although simultaneous observations of widely-separated pairs are used in a real application. Fair, unbiased switching depends on simultaneous control of local noise sources, which the dataset and analysis does incorporate, if only on minute-or-so timescales and longer, not the much-shorter cadence of a QM test. And even though no actual QM-test apparatus was implemented here, it does suggest this: employing widely-separated optical observatories can avoid the residual correlation sensed in these data, which is impossible for the constricted angular separations available to any single site. An attractive aspect of Gemini is that it provides a view stretching a full $180^{\circ}$ across the sky.

The Gemini GMOS-N and GMOS-S sample continues to grow, and although the current sample covers the full sky and range of possible angular separations between objects, there were not sufficient $z\geq3.65$ quasars to provide a meaningful comparison restricted to those; pursuing that is encouraged, and should be possible as more data are available. A potentially confusing factor in the current datset may be shifting bandpasses with redshift, and correcting to a common colour. Thus, a better technique may be spectroscopic, focused on bright emission lines. There was also limited information on how those individual quasars (or the calibration stars) may have varied during the sample period. A subset of the data with significant differences is one output of the work, and provides a baseline from which to compare. Repeat observations of these at higher photometric precision would seem to provide a check on either real, intrinsic correlation between those fluxes or false, spurious correlation due to an unidentified source-selection or instrumental effects. Those were well controlled here by the telescopes and instruments being essentially identical, and blind selection from a prior independent catalog, but a wealth of archival sources of photometry from other telescopes could be added together to improve on this result too. Future facilities allowing precision long-term monitoring, such as with the Vera C. Rubin Observatory, and now ongoing Large Synoptic Survey, will potentially provide an excellent database with which to do that. 



\vspace{6pt}

\funding{This research received no external funding.}

\dataavailability{All data discussed herein are freely available via public archives using the links indicated in the text, and their references. The spooky\_correlation code is available from \url{https://github.com/ericsteinbring/spooky_correlation}. It is written in IDL.
}

\acknowledgments{This research used the facilities of the Canadian Astronomy Data Centre operated by the National Research Council of Canada with the support of the Canadian Space Agency. Archival observations employed the Gemini Multi-Object Spectrgraphs (GMOS) on the North and South telescopes of the international Gemini Observatory, a program of NSF NOIRLab, which is managed by the Association of Universities for Research in Astronomy (AURA) under a cooperative agreement with the U.S. National Science Foundation on behalf of the Gemini partnership: the U.S. National Science Foundation (United States), National Research Council (Canada), Agencia Nacional de Investigación y Desarrollo (Chile), Ministerio de Ciencia, Tecnología e Innovación (Argentina), Ministério da Ciência, Tecnologia, Inovações e Comunicações (Brazil), and Korea Astronomy and Space Science Institute (Republic of Korea).}

\conflictsofinterest{The author declares no conflict of interest.} 





\appendixtitles{yes} 
\appendixstart
\appendix



\section{Outcomes of an Idealized Quasar Bell-Theorem Test}\label{appendix}

In the QM-test gedankenexperiment of \S{2.2}, one of two entangled electrons travels down each of two device arms to a rotatable waveplate and an observer (call them Alice and Bob).  Each arm also has a telescope that receives one of two external triggering photons, either red or blue from either quasars A or B. Alice and Bob can set their own waveplates, nominally to be either $0^\circ$ (red) or $180^\circ$ (blue) so each can make an electron-state detection (spin-down/spin-up). The test-result is the combined (average) outcome of the two arms, which in practice would be done collectively, repeated over many simultaneous trials.

\subsection{Truly Random External Photons}\label{appendix1}

For truly random red or blue photons, and because $D=2$ if detecting an electron to be spin-down (red) while $D=0$ whenever found spin-up (for either red or blue) there are only two distinct potential outcomes of the test, with equal (50\%) probability. If Alice receives a red photon, detecting spin-down (QM would imply the other electron state is spin-up) together Alice and Bob measure ${\bar D}=(2+0)/2=1$.  Otherwise, the outcomes for Alice and Bob are simply reversed when Alice instead receives a blue photon. Therefore, the probability from equation 2 is $\bar{P}=3/4$, or effectively reaching 75\% confidence, without influence by external sources. Exploiting the maxima of $D(\theta)$ can make ${\bar P}$ higher for better waveplate choices by Alice and Bob (as experimenters do in real tests), but that is not done here. This instead anticipates finding preferred angular separations for quasars A and B.

\subsection{Perfectly Correlated External Photons}\label{appendix2}

For a perfectly-correlated pair A and B, Alice and Bob can predict when the other will receive a red or blue photon, and set their own waveplate accordingly: ${\bar D}=(2+0)/2=1$. So whatever the separation of A and B, outcomes still mimic exactly the same QM test-result as before for random photons, but instead via external information (by definition, via a hidden variable). Notice, however, that when A and B are separated by $180^\circ$ on the sky this situation is also identically symmetric along the two arms (by Alice and Bob instead flipping their receiver-telescope orientation on the sky, i.e. swapping quasars A and B) so exactly reversed under reflection, with the {\it same} results each time. This is why the difference-measure $D(\phi)$ in equation 3 must be in quadrature, as it is soley a function of the great circle angle from A to B (or B to A) and necessarily continuous at $0^\circ$ and $180^\circ$. 



\reftitle{References}

\PublishersNote{}
\end{document}